\documentclass[aps,prl,preprint,tightenlines,superscriptaddress,showpacs,byrevtex]{revtex4}


\usepackage{notes2bib}
\usepackage{graphicx}
\usepackage{gensymb}
\usepackage{url}
\usepackage{graphicx}
\usepackage{amsmath}
\usepackage{latexsym}
\usepackage{amssymb}
\usepackage{color}

\usepackage{slashed}

\begin{document}

\preprint{\vbox{ \hbox{   }
                 \hbox{BELLE-CONF-1502}
}}




\title{\quad\\[0.5cm]Measurement of Azimuthal Modulations in the Cross-Section of Di-Pion Pairs in Di-Jet Production from Electron-Positron Annihilation}


\date{\today}
\noaffiliation
\affiliation{University of the Basque Country UPV/EHU, 48080 Bilbao}
\affiliation{Beihang University, Beijing 100191}
\affiliation{University of Bonn, 53115 Bonn}
\affiliation{Budker Institute of Nuclear Physics SB RAS and Novosibirsk State University, Novosibirsk 630090}
\affiliation{Faculty of Mathematics and Physics, Charles University, 121 16 Prague}
\affiliation{Chiba University, Chiba 263-8522}
\affiliation{Chonnam National University, Kwangju 660-701}
\affiliation{University of Cincinnati, Cincinnati, Ohio 45221}
\affiliation{Deutsches Elektronen--Synchrotron, 22607 Hamburg}
\affiliation{University of Florida, Gainesville, Florida 32611}
\affiliation{Department of Physics, Fu Jen Catholic University, Taipei 24205}
\affiliation{Justus-Liebig-Universit\"at Gie\ss{}en, 35392 Gie\ss{}en}
\affiliation{Gifu University, Gifu 501-1193}
\affiliation{II. Physikalisches Institut, Georg-August-Universit\"at G\"ottingen, 37073 G\"ottingen}
\affiliation{The Graduate University for Advanced Studies, Hayama 240-0193}
\affiliation{Gyeongsang National University, Chinju 660-701}
\affiliation{Hanyang University, Seoul 133-791}
\affiliation{University of Hawaii, Honolulu, Hawaii 96822}
\affiliation{High Energy Accelerator Research Organization (KEK), Tsukuba 305-0801}
\affiliation{Hiroshima Institute of Technology, Hiroshima 731-5193}
\affiliation{IKERBASQUE, Basque Foundation for Science, 48013 Bilbao}
\affiliation{University of Illinois at Urbana-Champaign, Urbana, Illinois 61801}
\affiliation{Indian Institute of Technology Bhubaneswar, Satya Nagar 751007}
\affiliation{Indian Institute of Technology Guwahati, Assam 781039}
\affiliation{Indian Institute of Technology Madras, Chennai 600036}
\affiliation{Indiana University, Bloomington, Indiana 47408}
\affiliation{Institute of High Energy Physics, Chinese Academy of Sciences, Beijing 100049}
\affiliation{Institute of High Energy Physics, Vienna 1050}
\affiliation{Institute for High Energy Physics, Protvino 142281}
\affiliation{Institute of Mathematical Sciences, Chennai 600113}
\affiliation{INFN - Sezione di Torino, 10125 Torino}
\affiliation{Institute for Theoretical and Experimental Physics, Moscow 117218}
\affiliation{J. Stefan Institute, 1000 Ljubljana}
\affiliation{Kanagawa University, Yokohama 221-8686}
\affiliation{Institut f\"ur Experimentelle Kernphysik, Karlsruher Institut f\"ur Technologie, 76131 Karlsruhe}
\affiliation{Kavli Institute for the Physics and Mathematics of the Universe (WPI), University of Tokyo, Kashiwa 277-8583}
\affiliation{King Abdulaziz City for Science and Technology, Riyadh 11442}
\affiliation{Department of Physics, Faculty of Science, King Abdulaziz University, Jeddah 21589}
\affiliation{Korea Institute of Science and Technology Information, Daejeon 305-806}
\affiliation{Korea University, Seoul 136-713}
\affiliation{Kyoto University, Kyoto 606-8502}
\affiliation{Kyungpook National University, Daegu 702-701}
\affiliation{\'Ecole Polytechnique F\'ed\'erale de Lausanne (EPFL), Lausanne 1015}
\affiliation{Faculty of Mathematics and Physics, University of Ljubljana, 1000 Ljubljana}
\affiliation{Ludwig Maximilians University, 80539 Munich}
\affiliation{Luther College, Decorah, Iowa 52101}
\affiliation{University of Maribor, 2000 Maribor}
\affiliation{Max-Planck-Institut f\"ur Physik, 80805 M\"unchen}
\affiliation{School of Physics, University of Melbourne, Victoria 3010}
\affiliation{Moscow Physical Engineering Institute, Moscow 115409}
\affiliation{Moscow Institute of Physics and Technology, Moscow Region 141700}
\affiliation{Graduate School of Science, Nagoya University, Nagoya 464-8602}
\affiliation{Kobayashi-Maskawa Institute, Nagoya University, Nagoya 464-8602}
\affiliation{Nara University of Education, Nara 630-8528}
\affiliation{Nara Women's University, Nara 630-8506}
\affiliation{National Central University, Chung-li 32054}
\affiliation{National United University, Miao Li 36003}
\affiliation{Department of Physics, National Taiwan University, Taipei 10617}
\affiliation{H. Niewodniczanski Institute of Nuclear Physics, Krakow 31-342}
\affiliation{Nippon Dental University, Niigata 951-8580}
\affiliation{Niigata University, Niigata 950-2181}
\affiliation{University of Nova Gorica, 5000 Nova Gorica}
\affiliation{Osaka City University, Osaka 558-8585}
\affiliation{Osaka University, Osaka 565-0871}
\affiliation{Pacific Northwest National Laboratory, Richland, Washington 99352}
\affiliation{Panjab University, Chandigarh 160014}
\affiliation{Peking University, Beijing 100871}
\affiliation{University of Pittsburgh, Pittsburgh, Pennsylvania 15260}
\affiliation{Punjab Agricultural University, Ludhiana 141004}
\affiliation{Research Center for Electron Photon Science, Tohoku University, Sendai 980-8578}
\affiliation{Research Center for Nuclear Physics, Osaka University, Osaka 567-0047}
\affiliation{RIKEN BNL Research Center, Upton, New York 11973}
\affiliation{Saga University, Saga 840-8502}
\affiliation{University of Science and Technology of China, Hefei 230026}
\affiliation{Seoul National University, Seoul 151-742}
\affiliation{Shinshu University, Nagano 390-8621}
\affiliation{Soongsil University, Seoul 156-743}
\affiliation{Sungkyunkwan University, Suwon 440-746}
\affiliation{School of Physics, University of Sydney, NSW 2006}
\affiliation{Department of Physics, Faculty of Science, University of Tabuk, Tabuk 71451}
\affiliation{Tata Institute of Fundamental Research, Mumbai 400005}
\affiliation{Excellence Cluster Universe, Technische Universit\"at M\"unchen, 85748 Garching}
\affiliation{Toho University, Funabashi 274-8510}
\affiliation{Tohoku Gakuin University, Tagajo 985-8537}
\affiliation{Tohoku University, Sendai 980-8578}
\affiliation{Earthquake Research Institute, University of Tokyo, Tokyo 113-0032}
\affiliation{Department of Physics, University of Tokyo, Tokyo 113-0033}
\affiliation{Tokyo Institute of Technology, Tokyo 152-8550}
\affiliation{Tokyo Metropolitan University, Tokyo 192-0397}
\affiliation{Tokyo University of Agriculture and Technology, Tokyo 184-8588}
\affiliation{University of Torino, 10124 Torino}
\affiliation{Toyama National College of Maritime Technology, Toyama 933-0293}
\affiliation{Utkal University, Bhubaneswar 751004}
\affiliation{CNP, Virginia Polytechnic Institute and State University, Blacksburg, Virginia 24061}
\affiliation{Wayne State University, Detroit, Michigan 48202}
\affiliation{Yamagata University, Yamagata 990-8560}
\affiliation{Yonsei University, Seoul 120-749}
  \author{A.~Abdesselam}\affiliation{Department of Physics, Faculty of Science, University of Tabuk, Tabuk 71451} 
  \author{I.~Adachi}\affiliation{High Energy Accelerator Research Organization (KEK), Tsukuba 305-0801}\affiliation{The Graduate University for Advanced Studies, Hayama 240-0193} 
  \author{K.~Adamczyk}\affiliation{H. Niewodniczanski Institute of Nuclear Physics, Krakow 31-342} 
  \author{H.~Aihara}\affiliation{Department of Physics, University of Tokyo, Tokyo 113-0033} 
  \author{S.~Al~Said}\affiliation{Department of Physics, Faculty of Science, University of Tabuk, Tabuk 71451}\affiliation{Department of Physics, Faculty of Science, King Abdulaziz University, Jeddah 21589} 
  \author{K.~Arinstein}\affiliation{Budker Institute of Nuclear Physics SB RAS and Novosibirsk State University, Novosibirsk 630090} 
  \author{Y.~Arita}\affiliation{Graduate School of Science, Nagoya University, Nagoya 464-8602} 
  \author{D.~M.~Asner}\affiliation{Pacific Northwest National Laboratory, Richland, Washington 99352} 
  \author{T.~Aso}\affiliation{Toyama National College of Maritime Technology, Toyama 933-0293} 
  \author{V.~Aulchenko}\affiliation{Budker Institute of Nuclear Physics SB RAS and Novosibirsk State University, Novosibirsk 630090} 
% ...
\collaboration{The Belle Collaboration}

\noaffiliation
\begin{abstract}
We present an extraction of azimuthal correlations between two pairs of charged pions detected in opposite jets from electron-positron annihilation.
These correlations may arise from the dependence of the di-pion  fragmentation on the polarization of the parent quark in the process $e^+e^- \rightarrow q \bar{q}$. Due to the correlation of the quark polarizations, the cross-section of di-pion pair production, in which the pion pairs  are detected in opposite jets in a dijet event, exhibits a modulation in the azimuthal angles of the planes containing the hadron pairs with respect to the production plane. 
The measurement of this modulation allows access to combinations of fragmentation functions that are sensitive to the quark's transverse polarization and helicity. Within our uncertainties we do not observe a significant signal from the previously unmeasured helicity dependent fragmentation function $G_1^\perp$.
This measurement uses a dataset of 938~fb$^{-1}$ collected by the Belle experiment  at or near $\sqrt{s}\approx10.58$ GeV.

\end{abstract}



\maketitle

\tighten

{\renewcommand{\thefootnote}{\fnsymbol{footnote}}}
\setcounter{footnote}{0}


Fragmentation functions are fundamental QCD objects describing the hadronization of quarks into final-state hadrons. They cannot be computed using lattice QCD techniques due to the  non-inclusive final-state and therefore have to be measured. 
In addition to studying properties of QCD, fragmentation functions are important tools to extract the partonic spin structure of the nucleon from semi-inclusive deep-inelastic scattering (SIDIS) and proton-proton (p+p) collision data~\cite{Brock:1993sz,deFlorian:2009vb}.
In recent years the role of polarization dependent fragmentation functions as quark polarimeters has become more important to access polarization dependent quark distributions in SIDIS and p+p~\cite{Barone:2001sp}.

Di-hadron fragmentation, e.g., \(q \to (h^{+}h^{-}) +X\), describes the
hadronization of a quark into a final-state containing two distinguishable hadrons. 
Compared to single-hadron fragmentation, it offers access to transverse
polarization of the fragmenting quark without involving either final-state
polarimetry or transverse-momentum dependence of the fragmentation
process. This is achieved through the additional degree of freedom
provided by the {\em relative} momentum of the hadron pair. In case of the
chiral-odd di-hadron fragmentation function \(H_{1}^{\sphericalangle}\), the
orientation of the plane spanned by this relative momentum and the total
di-hadron momentum can be correlated with the transverse polarization of
the fragmenting quark. Such correlation is reminiscent of the explicitly
transverse-momentum dependent Collins fragmentation function
\(H_{1}^{\perp}\), where the transverse momentum of the outgoing hadron
(or di-hadron) provides the plane correlated with the quark polarization.
However, such correlation and thus \(H_{1}^{\perp}\) does not survive
simple integration over transverse momentum, which complicates the
theoretical description. In contrast, the di-hadron plane does survive
such integration, permitting application of the ordinary collinear
framework.

The chiral-odd  \(H_{1}^{\sphericalangle}\) has already been subject to
measurements in \(e^{+}e^{-}\) annihilation into di-hadrons~\cite{Vossen:2011fk} as
well as in SIDIS and p+p. Its chiral-even partner, the helicity-dependent
di-hadron fragmentation function \(G_{1}^{\perp}\), has yet to be
measured. Together with the chiral-odd fragmentation functions and the
quark-polarization independent \(D_{1}\), it completes the set of
leading-twist di-hadron fragmentation functions.
Unlike \(H_{1}^{\sphericalangle}\), which has a counterpart in the single-hadron fragmentation with the Collins fragmentation function, for \(G_{1}^{\perp}\) no such correspondence exist~\cite{Boer:2003ya}. This makes it an elusive tool to probe quark helicity (e.g., in the nucleon) through fragmentation.
In addition, it has been conjectured that the
measurement of \(G_{1}^{\perp}\) in \(e^{+}e^{-} \to (h_{1}h_{2})
(\bar{h}_{1}\bar{h}_{2})+X \), i.e., production of two di-hadrons from
both the quark and the anti-quark in \(e^{+}e^{-}\) annihilation, is
related to jet-handedness correlations, which might receive contributions
from \(CP\) violating effects~\cite{Boer:2003ya}.

The process of di-hadron production from both the quark and the anti-quark
in \(e^{+}e^{-}\) annihilation is the cleanest way to access
polarization-dependent di-hadron fragmentation. While in SIDIS and p+p the
fragmentation functions are folded with partially unknown parton
distributions, such ``contamination''  is missing in \(e^{+}e^{-}\)
annihilation. Furthermore, the spin-correlation of the quark anti-quark
pair in \(e^{+}e^{-}\) annihilation allows the study of polarization
dependent fragmentation functions even in the case of an unpolarized
initial state. For the study of \(G_{1}^{\perp}\), \(e^{+}e^{-} \to
(h_{1}h_{2}) (\bar{h}_{1}\bar{h}_{2})+X \) has another advantage: formally
\(G_{1}^{\perp}\) is a transverse-momentum dependent fragmentation
function and as such does not survive simple integration over the
transverse momentum generated in the fragmentation process. However, the correlation between the two hadron pairs in this specific
process introduces a particular weighting factor in the integration, such
that the cross section itself survives integration over the {\em observed}
transverse momentum. The fragmentation functions will thus enter as
specific transverse-momentum moments in the cross section expressions.
However, we shall adopt the simplified notation of
\(G_{1}^{\perp}(z,M^{2})\) from Ref.~\cite{Boer:2003ya}, where \(z\) is the
fractional momentum carried by the hadron pair and \(M\) its invariant
mass.

This paper reports on the first measurement of the
\(G_{1}^{\perp}\)-induced correlation between two di-pions in the process
\(e^{+}e^{-} \to (\pi^{+}\pi^{-}) (\pi^{+}\pi^{-})+X \), where the two
di-pions are from two separate and opposing jets.

The Belle experiment~\cite{BelleDetector} at the KEKB storage ring~\cite{KEKB} recorded about 1~ab$^{-1}$ of $e^+e^-$ annihilation data. The data was taken mainly at the $\Upsilon(4S)$ resonance at $\sqrt{s}=10.58$~GeV, but also at $\Upsilon(1S)$ to $\Upsilon(5S)$ resonances and at $\sqrt{s}=10.52$~GeV.
The Belle instrumentation used in this analysis includes a central drift chamber and a silicon vertex detector, which provides precision tracking for tracks in $17.0\degree < \theta_\textrm{Lab}< 150.0\degree$, 
and electromagnetic calorimeters (ECL) covering the same region. 
Particle identification is performed using information on dE/dx in the CDC, a time-of-flight system in the barrel, aerogel Cherenkov counters in the barrel and the forward endcap, as well as a muon and $K_L$ identification system outside the superconducting solenoid, which provides a 1.5~T magnetic field.

Using the event shape variable thrust, a sample of light and charm quark ($u,d,s,c$) fragmentation can be selected~\cite{Seidl:2008xc, Vossen:2011fk}. 
The thrust is defined as 
\begin{equation}
T= \max_{|\vec{n}|=1} \frac{\sum_i |\vec{n}\cdot \vec{p}_i|}{\sum_i |\vec{p}_i|}.
\end{equation}
Here the $\vec{p}_i$ are all particles in the event. We use a thrust cut of $T>0.8$ which reduces the contribution of $B$ mesons to less than 1\%. 
Since the polarization-dependence of the fragmentation would average out in unpolarized $e^+e^-$ annihilation into one di-pion, a measurement has to use the correlation between the polarizations of the produced quark-antiquark pairs.  This then leads to a correlation of hadron-pairs detected in opposite hemispheres~\cite{Boer:2003ya}. 
The correlation is expressed as a modulation of the di-hadron pair cross-section in the angles $\Phi_{R^\prime_1}$ and $\Phi_{R^\prime_2}$. They are defined in Ref.~\cite{Boer:2003ya} as the azimuthal angles of the planes containing the hadron pairs with respect to the event plane spanned by the beam axis and the momentum sum vector $\vec{P}_h=\vec{h}_1+\vec{h}_2$ of one hadron pair in the Center-of-Mass System(CMS), where $h_1$, $h_2$ are the momenta of the hadrons forming the pair. It should be noted, that the reference frame defined here, is {\em asymmetric} with respect to the hadron pairs, both angles $\Phi_{R^\prime_1}$ and $\Phi_{R^\prime_2}$ are measured around $\vec{P}_h$. With this definition, the di-pion fragmentation function $G_1^\perp$ can be accessed in the moment

\begin{equation}
\left< \cos( 2( \Phi_{R^\prime_1}- \Phi_{R^\prime_2} ) )  \right> \propto \sum_{q,\bar{q} } e^2_q  G_1^{\perp,q} (z,M^2) G_1^{\perp,\bar{q}}(\bar{z},\bar{M}^2).
\label{eq:g1t}
\end{equation}
Here $q$, $\bar{q}$ are the quark and antiquark flavors, $e_q$ the corresponding electric charges and $\bar{z}$, $\bar{M}$ the fractional energy and invariant mass of the hadron pair produced in the antiquark fragmentation.  

\begin{figure}
\includegraphics[width=0.95\textwidth]{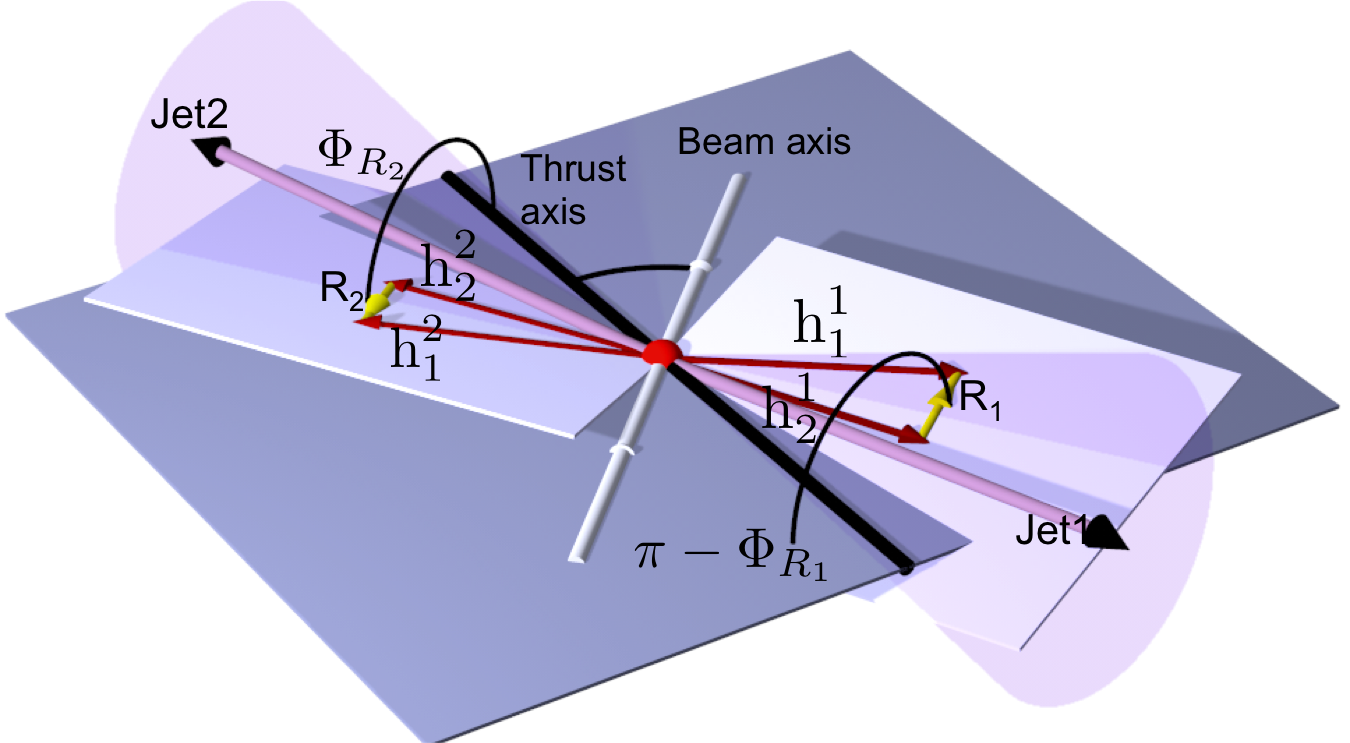}
\caption{The coordinate system used for this measurement. For convenience, the jets are depicted to lie in the event plane which in turn is spanned by the beam axis and the jets.\label{fig:coo}}
\end{figure}

Instead of using the {\em asymmetric} frame defined above, this analysis uses a {\em symmetric} frame in which the angles are computed using the jet axes of di-jet events. The jet axes are reconstructed using the anti-$k_T$ jet algorithm~\cite{Cacciari:2008gp}. They serve here as proxies of the directions of the fragmenting (anti)quarks.
A similar approach had been employed in the previous measurements of $H_1^\sphericalangle$ and the Collins fragmentation function~\cite{Vossen:2011fk,Seidl:2008xc} in a {\em symmetric} frame using the thrust axis.  
The coordinate system is depicted in Fig.~\ref{fig:coo}.

The angles $\Phi_{R_i}$ can be expressed as:

\begin{equation}
\Phi_{R_i}=\textrm{sgn}\left(\vec{j_i}\cdot\left((\vec{e}\times \vec{j_i})\times(\vec{j_i}\times\vec{R_i}) \right) \right) \arccos\left( \frac{\vec{e}\times \vec{j_i}}{|\vec{e}\times\vec{j_i}|} \cdot \frac{\vec{j_i} \times \vec{R_i}}{|\vec{j_i}\times\vec{R_i}|}  \right).
\label{eq:phiR}
\end{equation}
Here $\vec{e}$ is the $e^-$-beam direction, $\vec{j_i}$ the jet axes and $\vec{R_i}=\vec{h}^i_1-\vec{h}^i_2$, with $\vec{h}^i_1$ the momentum of the positive pion and $\vec{h}^i_2$ the momentum of the negative pion in jet $i$. All quantities are in the CMS. Note that one of the jet-axis directions is flipped such that in an ideal back-to-back event, both jet axes point in the same direction. 
Compared to the asymmetric frame, the theory for the symmetric frame has not been worked out completely, though we believe it to be just a modification of Eq.~(\ref{eq:g1t}).
Compared to~\cite{Vossen:2011fk} and~\cite{Seidl:2008xc}, where the thrust axis of the event was used as a proxy for the quark direction, the jet method will be more robust against final-state gluon radiation, which is usually described by Sudakov factors or evolution equations.

We use jets reconstructed with the anti-$k_T$ algorithm as implemented in the FastJet 3.0.5 package~\cite{Cacciari:2011ma,hep-ph/0512210} with a jet radius of $R=1.0$ and a minimum energy of 3.75~GeV for each jet. This cut on the jet energy also effectively removes $e^+e^- \rightarrow \tau \bar{\tau}$ events and leaves only dijet events.
We restrict the allowed range of the polar angle of the jets in the CMS to $1.38~\textrm{rad} < \theta_\textrm{CMS}< 1.75~\textrm{rad}$ to minimize systematic effects at the edges of the barrel acceptance. Jets are reconstructed from all charged tracks that originate from a region around the reconstructed interaction point (IP) as well as photons with energies larger than 50~MeV in the barrel, 75~MeV in the forward endcap and 100~MeV in the backward endcap. The IP is required to fulfill $|dr| < 2$~cm and $|dz|< 4$~cm, where $dr$ and $dz$ are the distance of closest approach to the interaction point in the plane perpendicular to the beam direction and along the direction of the beams.
We select charged pion pairs by using all charged tracks, which have not been identified as kaons, protons, electrons, or muons~\cite{Nakano2002402} and requiring a minimum fractional energy $z$ of the individual tracks of 0.1. 

\begin{figure}
\includegraphics[width=0.48\textwidth]{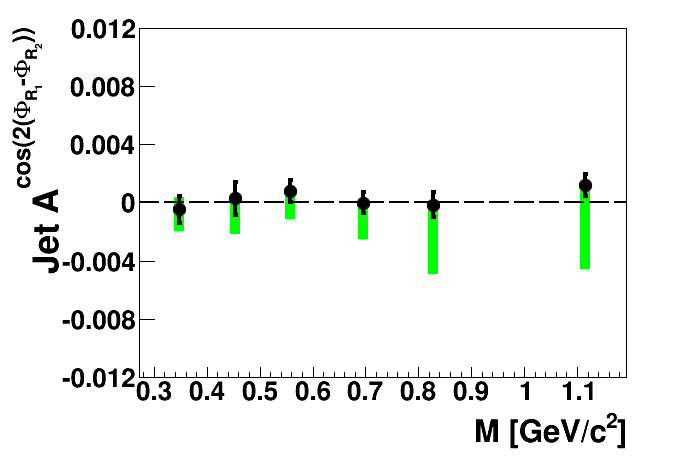}\includegraphics[width=0.48\textwidth]{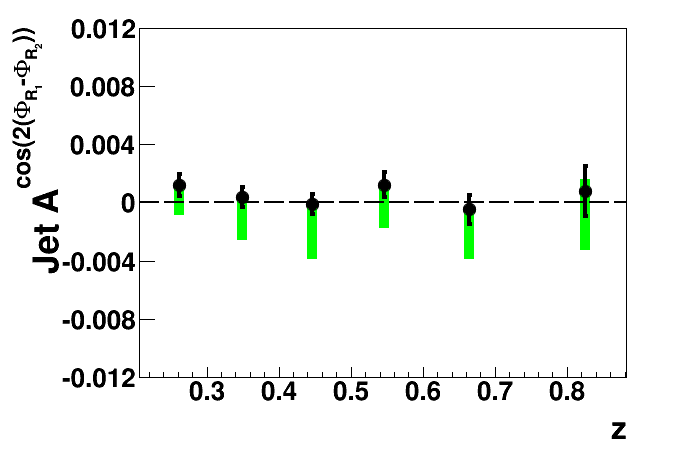}
\caption{Results for $A^{\cos\left(2(\Phi_{R_1}-\Phi_{R_2})\right)}$ binned in $M$ and $z$.\label{fig:g1t} The black error bars are statistical and the green  bands show the systematic uncertainty.}
\end{figure}

We perform a two-dimensional $\chi^2$ fit to the normalized di-pion pair yields  $\frac{N(\Phi_{R_1},\Phi_{R_2})}{\left<N\right>}$ 
using the truncated Fourier expansion
\begin{equation}
1+A^{\cos\left(\Phi_{R_1}+\Phi_{R_2}\right)}\cos\left(\Phi_{R_1}+\Phi_{R_2}\right)+A^{\cos\left(2(\Phi_{R_1}-\Phi_{R_2})\right)}\cos\left(2(\Phi_{R_1}-\Phi_{R_2})\right).
\end{equation}
The amplitude $A^{\cos\left(2(\Phi_{R_1}-\Phi_{R_2})\right)}$ corresponds then to the quantities in Eq.~(\ref{eq:g1t}), whereas the amplitude $A^{\cos\left(\Phi_{R_1}+\Phi_{R_2}\right)}$ corresponds to the previously published measurement of $H_1^\sphericalangle$~\cite{Vossen:2011fk}.
Results binned in $M$ and $z$ of one hadron pair are shown in Fig.~\ref{fig:g1t} integrated over $z$ and $M$ of the other pair. 

The purity of the sample is above 90\% for di-pion pairs, with the remaining 10\% coming from an admixture of kaons. The leading systematic effects are the contributions from detector smearing and fiducial cut edge effects to the measured asymmetries. These are estimated from simulations using Pythia~\cite{Sjostrand:2006za} and EvtGen~\cite{Lange:2001uf} for various physics processes not including the polarization dependent ones of Eq.~(\ref{eq:g1t}), and GEANT3~\cite{Brun:1987ma} for the detector effects.  
In addition to the light ($u,d,s$) quark fragmentation, our results also contain significant contributions from charm quark production. The charm contribution is 18\% on average and approximately constant over the invariant mass bins. In $z$, it drops monotonically from 21\% to 5\%.

Within our statistical and systematic uncertainties, we do not observe a deviation from zero for $A^{\cos\left(2(\Phi_{R_1}-\Phi_{R_2})\right)}$. Statistical uncertainties are symmetric and about $10^{-3}$ in magnitude, whereas our systematic uncertainty is asymmetric and varies between $2\cdot 10^{-3}$ at low $M$ and low $z$ and $5\cdot 10^{-3}$ at high $M$ and $z$.
It has been conjectured that the $G_1^\perp$
	fragmentation function receives non-zero contributions mainly from
	$p$-$p$ wave interference in the partial-wave expansion of $G_1^\perp$. We
	note that an enhancement of such contribution by restricting
	$\theta_\text{decay}$, the decay angle of the positive hadron in
	the hadron-pair CMS used in such expansion, to values for which
	$\cos(\theta_\text{decay})$ is positive does not result in a
	significantly non-zero amplitude either.


In summary, we show first results for azimuthal modulations in the cross-section of di-pion pairs in di-jet production from electron-positron annihilation. 
The amplitude of the $\cos\left(2(\Phi_{R_1}-\Phi_{R_2})\right)$ modulation, which is sensitive to the helicity dependent fragmentation function $G_1^\perp$, is consistent with zero within our statistical and systematic uncertainties. 

We thank the KEKB group for the excellent operation of the
accelerator; the KEK cryogenics group for the efficient
operation of the solenoid; and the KEK computer group,
the National Institute of Informatics, and the 
PNNL/EMSL computing group for valuable computing
and SINET4 network support.  We acknowledge support from
the Ministry of Education, Culture, Sports, Science, and
Technology (MEXT) of Japan, the Japan Society for the 
Promotion of Science (JSPS), and the Tau-Lepton Physics 
Research Center of Nagoya University; 
the Australian Research Council and the Australian 
Department of Industry, Innovation, Science and Research;
Austrian Science Fund under Grant No.~P 22742-N16 and P 26794-N20;
the National Natural Science Foundation of China under Contracts 
No.~10575109, No.~10775142, No.~10875115, No.~11175187, and  No.~11475187; 
the Ministry of Education, Youth and Sports of the Czech
Republic under Contract No.~LG14034;
the Carl Zeiss Foundation, the Deutsche Forschungsgemeinschaft
and the VolkswagenStiftung;
the Department of Science and Technology of India; 
the Istituto Nazionale di Fisica Nucleare of Italy; 
National Research Foundation (NRF) of Korea Grants
No.~2011-0029457, No.~2012-0008143, No.~2012R1A1A2008330, 
No.~2013R1A1A3007772, No.~2014R1A2A2A01005286, No.~2014R1A2A2A01002734, 
No.~2014R1A1A2006456;
the Basic Research Lab program under NRF Grant No.~KRF-2011-0020333, 
No.~KRF-2011-0021196, Center for Korean J-PARC Users, No.~NRF-2013K1A3A7A06056592; 
the Brain Korea 21-Plus program and the Global Science Experimental Data 
Hub Center of the Korea Institute of Science and Technology Information;
the Polish Ministry of Science and Higher Education and 
the National Science Center;
the Ministry of Education and Science of the Russian Federation and
the Russian Foundation for Basic Research;
the Slovenian Research Agency;
the Basque Foundation for Science (IKERBASQUE) and 
the Euskal Herriko Unibertsitatea (UPV/EHU) under program UFI 11/55 (Spain);
the Swiss National Science Foundation; the National Science Council
and the Ministry of Education of Taiwan; and the U.S.\
Department of Energy and the National Science Foundation.
This work is supported by a Grant-in-Aid from MEXT for 
Science Research in a Priority Area (``New Development of 
Flavor Physics'') and from JSPS for Creative Scientific 
Research (``Evolution of Tau-lepton Physics'').





\bibliography{BelleDiHadModPaper}


\end{document}